\documentclass{article}
\usepackage{amsfonts,amssymb,latexsym,amscd}
\begin{document}

\title{ \sc Evolution of the equation of state parameters of cosmological tachyonic  field components through mutual interaction}
\author{Murli Manohar Verma$^{*}$, Shankar Dayal Pathak$^{**}$\\
{\small   Department of Physics, Lucknow University, Lucknow  226 007, India} \\
{  *sunilmmv@yahoo.com}\\
{ ** prince.pathak19@gmail.com}
}

\date{}
\maketitle

\begin{abstract}
We study  the perturbed equation of state (EOS) parameters of the cosmological tachyonic scalar field  components and their mutual time-dependent interaction.
 It is shown that the discrete temperature-dependent pattern of the EOS emerges from an initial continuum along the evolution of the universe. This leads to
 two major components in form of dark energy and dark matter, and also suggests a solution to the cosmological constant problem and the coincidence problem.
\end{abstract}

\section{\textbf{Introduction}}
   In the backdrop of the  work done in the standard  cosmological model for quintessence dark energy and dark matter \cite{a1}, we propose the  evolutionary  history of the universe with interacting dark energy--- in form of a part  of the string-inspired  tachyonic scalar field \cite{a2}.  The other component mimics  the  cosmological constant of inflation era.  However,  each component undergoes  a perturbation and so changes flavours.  The mutual interaction makes the cosmological constant drop from a very high value to the very small one, and thus leads to a possible solution of  the cosmological constant problem \cite {a3}.  The EOS of the field depends on temperature,   condensing   into a discrete characteristic  pattern from a continuum as the universe expands and cools. It is, therefore,  reasonably  expected  that this  condensation must  grow faster with the ongoing acceleration of the universe. These aspects can be  constrained  with the current observations of the Hubble parameter at various redshifts.

\section{\textbf {Perturbed equation of state parameters}}

 A small perturbation $\varepsilon(t)$ in the  EOS of the first component of the tachyon field---the cosmological constant---  now called  shifted cosmological parameter(SCP), is introduced    with  the new EOS  as $w'_{\lambda}=-1+\varepsilon(t)$.  The  energy density  and pressure of SCP respectively are  $\rho'_{\lambda}=V(\phi)\sqrt{1-\partial^{i}\phi\partial_{i}\phi}$ and $ P'_{\lambda}=-V(\phi)\sqrt{1-\partial^{i}\phi\partial_{i}\phi} + \varepsilon V(\phi)\sqrt{1-\partial^{i}\phi\partial_{i}\phi}$.  As the second component, shifted dust matter (SDM)  has pressure $ P'_{m}=-\varepsilon V(\phi)\sqrt{1-\partial^{i}\phi\partial_{i}\phi} $  with  EOS $ w'_{m}= -\frac{\varepsilon}{\partial^{i}\phi\partial_{i}\phi}+\varepsilon $,
while the EOS of the over-all  tachyonic scalar field is $ w_{total}=(\partial^{i}\phi\partial_{i}\phi-1)$.

Considering the time-dependent interaction strength among the field components,   $ Q=\gamma\dot{\rho'}_{m}$   and  $\dot{\phi}^{2}<<V(\phi)$,  we have SDM energy density scaling as
\begin{eqnarray}\rho'_{m}=\rho'^{0}_{m}\left(\frac{a}{a_{0}}\right)^{-\frac{3}{1-\gamma}(1+\varepsilon-\varepsilon/\dot{\phi}^{2})}.\label{n16}\end{eqnarray}

It is found  that  $w'_{m}+w'_{\lambda}=-1-\varepsilon/\dot{\phi}^{2}+2\varepsilon$,  that is,   $w'_{m}+w'_{\lambda}\neq w_{total}$
while $w'_{m}+w'_{\lambda}= w_{total} + w'$ where $w'=-\varepsilon/\dot{\phi}^{2}-\dot{\phi}^{2}+2\varepsilon$. For SCP

 \begin{eqnarray}\rho'_{\lambda}=\rho'^{0}_{\lambda}x^{3\varepsilon} -\frac{\gamma\rho'^{0}_{m}(1-\varepsilon/\dot{\phi}^{2}+\varepsilon)}{1-\varepsilon/\dot{\phi}^{2}+\gamma\varepsilon}\left[x^{\frac{3}{1-\gamma}(1-\varepsilon/\dot{\phi}^{2}+\varepsilon)}-x^{3\varepsilon}\right]\label{n18}\end{eqnarray} where $x=a_{0}/a=1+z$, and  in the absence of perturbation  ($\varepsilon\rightarrow 0$) we have $\rho'_{\lambda}\rightarrow\rho'^{0}_{\lambda}-\gamma\rho'^{0}_{m}\left[(\frac{a_{0}}{a})^{3/1-\gamma}-1\right]$.

The  Friedmann equation  for spatially flat universe ($k=0$) now becomes

\begin{eqnarray}H^{2}(x)=H^{2}_{0}\left[\Omega'^{0}_{m}x^{\alpha}+\Omega'^{0}_{\lambda}x^{3\varepsilon}+\frac{\Omega'^{0}_{m}\gamma(1+\varepsilon-\varepsilon/\dot{\phi}^{2})}{(1+\gamma\varepsilon-\varepsilon/\dot{\phi}^{2})}\left(x^{3\varepsilon}-x^{\alpha}\right)\right].\label{m13}\end{eqnarray}

We can also  determine the epoch of equality $(z_{eq})$  at which $\rho'_{m}=\rho'_{\lambda}$ by solving

\begin{eqnarray}(1+z_{eq})^{\alpha-3\varepsilon}+\frac{\gamma(1-\varepsilon/\dot{\phi}^{2}+\varepsilon)}{(1-\varepsilon/\dot{\phi}^{2}+\gamma\varepsilon)}\left[(1+z_{eq})^{\alpha-3\varepsilon}-1\right]=\frac{\Omega'^{0}_{\lambda}}{\Omega'^{0}_{m}}\label{n19}\end{eqnarray} where $\alpha =\frac{3}{1-\gamma}(1+\varepsilon-\varepsilon/\dot{\phi}^{2})$ and $\Omega'^{0}_{\lambda}$  and  $\Omega'^{0}_{m}$ are the present values of density parameter of the cosmological constant and matter respectively. Taking  $\Omega'^{0}_{\lambda}=0.73$ and $\Omega'^{0}_{m}=0.27$  we have ${\Omega'^{0}_{\lambda}}/{\Omega'^{0}_{m}}\approx 2.70$. The values  of $\gamma$ and $\varepsilon$ can be inferred  from observations.  In the absence of perturbation and interaction $z_{eq}\simeq0.3932$.  At specific  $z=z'$ when the values  of $\dot{\rho'_{\lambda}}$ and $\dot{\rho'_{m}}$ are equal,   we get
\begin{eqnarray}\frac{\rho'_{m}}{\rho'_{\lambda}}=\frac{\varepsilon (1-\gamma)}{(1+\gamma)(1-\varepsilon/\dot{\phi}^{2}+\varepsilon)}.\label{n20}\end{eqnarray} Clearly,  from (\ref{n20}) we see that in the absence of perturbation $(\varepsilon\rightarrow0)$ the ratio ${\rho'_{m}}/{\rho'_{\lambda}}\rightarrow 0$,  which is un-physical.

\section{Temperature-dependence of the EOS}

 The tachyonic field EOS $w_{\phi} =\dot{\phi}^{2}-1$ crucially hinges on  the kinetic energy term   which is naturally expected to be  function of temperature. With the effective  EOS  as   $ \tilde{w}(T)=\dot{\phi}^{2}/2$, the field  energy density evolves as

\begin{equation}\label{n8}
   \rho_{\phi}=\rho^{0}_{\phi}\left(\frac{a}{a_{0}}\right)^{-6\tilde{w}(T)}.
\end{equation}

Consequent to  the  temperature variation  $T\propto a^{-1}$, the  EOS of cosmic  tachyonic  scalar field changes accordingly ($\tilde{w}(T)=0$, $1/2$ or $2/3$  correspond to the cosmological constant, pressureless  matter  or   radiation  respectively).
 The break-up  of the continuum  of EOS  into a  discrete temperature-dependent  pattern   manifesting in distinct forms of components   resembles the  symmetry breaking phenomena  in the  early universe.
Although the present observations do  not unambiguously constrain the exact form of kinetic energy of the tachyonic  field as a  function of temperature,  we study  two forms,\textit{ viz}.,  $ \dot{\phi}^{2}(T)=\alpha T^{n} exp(\beta T^{m})$   and $ \dot{\phi}^{2}(T)=A T^{n} + B exp(\beta T^{m})$   where $\alpha, \beta$ and $A, B$ are constants.

\section{\textbf{Conclusion }} We have shown that the  interaction strength among tachyonic field components with perturbed EOS ascertains their respective evolution. The interaction strength itself  can be constrained by the Hubble parameter observations at different redshifts. The interaction also  alleviates  the cosmological constant problem.  The EOS of the field evolves and condenses, as the universe cools,    from a continuum to discrete pattern representing the forms---  dark energy and dark matter--- which  we mostly observe in the universe at present. The break-up of these  compatible forms at a  specific epoch indicates a possible explanation of the coincidence problem.

\section{\textbf{Acknowledgement}} The  authors  acknowledge  the financial support F. No. 37-431/2009 (SR) received from UGC, New Delhi for   pursuance of  this work.

\end{document}